\documentclass[reprint,amsmath,amssymb,aps]{revtex4-2}
\usepackage{graphicx}% Include figure files
\usepackage{dcolumn}% 
\usepackage{bm}% bold math
\usepackage{mathrsfs}
\usepackage{amsmath}
\usepackage{hyperref}
\usepackage{amssymb}
\usepackage[dvipsnames]{xcolor}

\newcommand{\dd}{\mathnormal{d}}
\newcommand{\old}[1]{}

\begin{document}

\preprint{APS/123-QED}
\title{Thin accretion disk around the distorted Schwarzschild black hole}% Force line breaks %with \\

\author{Shokoufe Faraji}
\email{shokoufe.faraji@zarm.uni-bremen.de}
 %\altaffiliation{University of Bremen, Center of Applied Space Technology and Microgravity(ZARM), 28359 Bremen}%Lines break automatically or can be forced with \\
\author{Eva Hackmann}%
 \email{eva.hackmann@zarm.uni-bremen.de}
\affiliation{%
University of Bremen, Center of Applied Space Technology and Microgravity(ZARM), 28359 Bremen}%
\begin{abstract}
We construct the relativistic standard steady, optically thick, cold and geometrically thin accretion disk around a distorted Schwarzschild black hole. The distortion of this static and axially
symmetric black hole solution, is connected to an external distribution of matter. If the effects due to the rotation are negligible, this distribution can describe the exterior of any axially symmetric astrophysical model. Also, the distortion could be related to the outer parts of the accretion disk. We study the effects due to a distortion up to the quadrupole, and compare the physical characteristics
of this disk to the usual Schwarzschild case.

\end{abstract}

\maketitle

\section{Introduction}
Several important astrophysical phenomena, such as active galactic nuclei or X-ray binaries, are powered by accretion onto black holes. The accretion of matter is one of the most powerful energy sources in the universe. 
 
% This area of studying accretion onto a compact object is always in the major interest of the astrophysics and have been vastly studied in the literature. Beside the sophisticated numerical simulations, also a several analytical accretion disk models have been vastly studied in the literature, for a review see e.g. \cite{Abramowicz2013}.

The study of accretion onto compact objects is therefore a highly interesting topic that has been discussed extensively in the literature. An important approach to this, in addition to sophisticated numerical simulations, is to find the analytical solution to accretion disk models. This approach has been in many aspects essential to understand the fundamental physical processes of accretion. For a review of accretion disk models see e.g.~\cite{Abramowicz2013}.

The standard thin accretion disk model is applied for a sub-critical accretion rate $M\leqslant M_{\rm{Edd}}$, where the critical accretion rate is defined by $M_{\rm{Edd}}(=16L_{\rm{Edd}}c^2)$, and $L_{\rm{Edd}}$ is the Eddington luminosity. This model was introduced in one of the first studies of accretion disks around black holes by Shakura \& Sunyaev in 1972 \cite{1973A&A....24..337S}. They presented the basic assumptions and equations to describe classic thin disks in the Newtonian regime. Relativistic disks were first described by Bardeen, Press \& Teukolsky in 1972 \cite{1972ApJ...178..347B}, and Novikov \& Thorne in 1973 \cite{1973blho.conf..343N}. Later, Paczynski \& Bisnovatyi-Kogan \cite{1981AcA....31..283P}, Muchotrzeb \& Paczynski  \cite{1982AcA....32....1M}, and Abramowicz et al \cite{1988ApJ...332..646A}, introduced a model named by latter slim disk, which includes a number of terms neglected in the thin disk assumptions. Besides, twenty years after introducing the standard thin disk, Riffert \& Herold corrected an algebraic error in the Novikov and Thorne model in the vertical structure equation \cite{1995ApJ...450..508R}.

The above mentioned papers considered the accretion onto a black hole described by the Kerr metric. Within the context of general relativity, models of accretion disks were generalised to different space-times. Stuchl{\'{\i}}k \& Slan{\'y} added a cosmological constant \cite{Stuchlik2004}, and Chen \& Jing \cite{Chen2012} analysed thin accretion disks in the Johannsen-Psaltis parametrised space-time. Properties of accretion disks 
around naked singularities have been studied in \cite{Kovacs2010}. More exotic objects were also considered, including Boson stars \cite{Torres2002}, wormhole space-times \cite{Harko2009}, quark stars \cite{Kovacs2009}, gravastars \cite{Harko2009c}, and Bose-Einstein condensate stars \cite{Danila2015}. Extensions of the thin disk model to some alternative theories of gravity have also been achieved. This includes $f(R)$ theories \cite{Pun2008,Perez2013}, effects of massless scalar fields \cite{Gyulchev2019}, Ho{\v{r}}ava gravity \cite{Harko2011}, brane world scenarios \cite{Heydari-Fard2010}, Chern-Simons modified gravity \cite{Harko2010}, scalar-vector-tensor theory \cite{Perez2017}, an higher dimensional Kaluza-Klein black hole \cite{Chen2012}, and a heterotic string theory model \cite{Karimov2018}.

All these models for the central compact object, both within general relativity and alternative theories of gravity, assume that the space-time outside the black hole does not contain any additional matter. It is however clear that in a real astrophysical environment the black hole is not isolated, but in particular surrounded by the accretion disk and also by other additional matter, radiation, or electromagnetic fields. Usually it is assumed, both in accretion disk simulations and in analytical accretion disk models, the surrounding energy content is in such a way that its influence on the space-time geometry can be neglected. The gravitational field produced by the disk itself may however have an important effect in active galactic nuclei, in particular for the outer parts of the disk \cite{Lodato2007}. The vast literature exist on the construction of solutions for black hole-disk systems, of which we cite here only the classical papers by Will \cite{Will1974} and Lemos \& Letelier \cite{Lemos1994}.

In this work we concentrate on a particular class of black holes within general relativity, the Schwarzschild black hole in the presence of external axially symmetric distribution of mass outside the horizon. For this we use a local solution of the Einstein equations. This simplifies to the Schwarzschild solution if the external matter is vanished. This space-time in the vicinity of horizon remain vacuum with the cost of relaxing assumption of asymptotically flatness in the static vaccum Einstein field equation. 

The first to consider the Schwarzschild black hole in an external gravitational field up to a quadrupole were Doroshkevich et al in 1965. In addition to construct the metric, they showed the regularity of the horizon in this case \cite{1965ZhETF...49.170D}. A detailed analysis of the global properties of the distorted Schwarzschild space-time was presented in the paper by Geroch \& Hartle \cite{1982JMP....23..680G}. In 2002, Chandrasekhar introduced the equilibrium condition for the black hole in a static external gravitational field in terms of the multiple moments \cite{Chandrasekhar:579245}. Geodesic motion in a distorted Schwarzschild black hole has been studied e.g. in \cite{2016PhRvD..93f4019S,Grover2018}.

In this paper we discuss the construction of the standard relativistic thin disk around a distorted Schwarzschild black hole up to the quadrupole. One may consider the distortion due to the external mass distribution as the contribution of the outer part of the disk, where self-gravity effects are expected to play a role. %to the geometry of the central black hole. 
Then using the distorted geometry, one can construct inner part of the standard thin accretion disk, to see the impact on the innermost parts of the disk. As an example, a recent study \cite{2016PhRvD..93f4019S} suggests that the innermost stable circular orbit, assumed to mark the inner edge of the disk, could be influenced significantly by the distortion of the black hole.

We adopted the convention $c=1$ and $G=1$, unless for section \ref{sec:results}. If somewhere we used other units we specified there. %Also, we used  Schwarzschild coordinate and Weyl coordinate for representing the distorted Schwarschild metric. In calculation $\sigma_{x\phi}$, the shear stress, we used local rest frame (LRF) of fluid as usual in standard thin disk model.

The plan of this paper is as follows. In section \ref{sec:metric} we explain the distorted Schwarzschild black hole. Then we review in  section \ref{sec:disk} the relativistic thin disk model, and we discuss the thin disk assumptions and the basic dynamical equations. In section \ref{sec:eq} we construct the thin disk around the distorted Schwarzschild black hole. We present the solution and results in section \ref{sec:results}. Finally, we summarise and conclude in section \ref{sec:discuss}.
%........................................222............................................................

\section{Distorted Schwarzschild black hole}\label{sec:metric}
%Uniqueness theorem is one of the important results in general relativity. For example in the case of Schwarzschild black hole, 
The Schwarzschild space-time is the unique static solution of Einstein's vacuum equation which is asymptotically flat with a smooth event horizon. However, there is a static vacuum solution with a regular event horizon, which is not asymptotically flat. This solution is obtained by assuming the existence of an external distribution of matter in such a way that the black hole is no longer isolated, and this assumption relaxes the asymptotic flatness, see e.g. \cite{Chandrasekhar:579245}.

Astrophysical black holes are assumed to rotate. Here, as a first step, we assume for simplicity that the rotation is so slow that it can be neglected. Therefore, the exterior of the black hole, in the presence of an external static and axially symmetric matter distribution, locally can be described by the distorted Schwarzschild black hole. 
% Most astrophysical models are studied, considered axially symmetric objects and by neglecting their rotation, exterior of them can describe by this solution. Weyl metric is commonly used to construct such a solution.
This approach was introduced first by Geroch and Hartle \cite{1982JMP....23..680G}. The Weyl metric is given by
\begin{align}\label{GHmetric}
\dd s^2 & =-e^{2\psi}\dd t^2 +e^{2(\gamma-\psi)}(\dd\rho^2+\dd z^2) +e^{-2\psi}\rho^2 \dd\phi^2\,.
\end{align}
The relation between Schwarzschild coordinates and Weyl coordinates is given by $\rho=\sqrt{r(r-2M)}\sin{\theta}$ and $z=(r-M)\cos{\theta}$. The Weyl metric contains two functions $\psi$ and $\gamma$, which only depend on $\rho$ and $z$. For the Schwarzschild solution, in the Schwarzschild coordinates, the metric functions are given by
\begin{align}
\psi_s&=\frac{1}{2}\ln\left(1-\frac{2M}{r}\right),\\ \gamma_s&=\frac{1}{2}\ln\left(1+\frac{M^2\sin^2\theta}{r^2-2Mr}\right).
\end{align}

The function $\psi$ plays the role of a  gravitational potential and obeys the Laplace equation in flat space $ds^2=d\rho^2+dz^2+\rho^2 d\phi^2$, meaning
\begin{align}
\psi_{s_,\rho\rho}+\frac{1}{\rho}\psi_{s_,\rho}+\psi_{s_,zz}=0.\label{laplap}
\end{align}
%\textcolor{red}{Why is here $\psi_s$ in the equation? Shouldn't it be just $\psi$?} sh: NO!!!!

It is convenient to rewrite the metric in prolate spheroidal coordinate $(t, x, y, \phi)$. The relation of this coordinates to the Weyl coordinates are as follows
\begin{align}
    {x} & =\frac{1}{2{M}}(\sqrt{\rho^2+({z}+{M})^2}+\sqrt{\rho^2+({z}-{M})^2})\,,\\
    {y} & =\frac{1}{2{M}}(\sqrt{\rho^2+({z}+{M})^2}-\sqrt{\rho^2+({z}-{M})^2}),\\
    {t}&=t,\\
     {\phi}&=\phi.
\end{align}
Equivalently, $\rho={M}\sqrt{({x}^2-1)(1-{y}^2)}$ and  ${z}={M}{x}{y}$. Here $M$ is a parameter which can be identified as the mass of the black hole. In these coordinates, the Schwarzschild solutions $\psi_s$ and $\gamma_s$ read as
\begin{align}
     \psi_s & =\frac{1}{2}\ln{\frac{{x}-1}{{x}+1}}\,,\\
     \gamma_s & =\frac{1}{2}\ln{\frac{{x}^2-1}{{x}^2-{y}^2}}\,.
\end{align}
Ultimately, the metric \eqref{GHmetric} takes the form
\begin{align}\label{Dmetric}
	\dd s^2 = &- \left( \frac{x-1}{x+1} \right) e^{2\hat{\psi}} \dd t^2+ M^2 (x+1)^2 e^{-2\hat{\psi}} \\
	& \left[e^{2\hat{\gamma}} \left( \frac{\dd x^2}{x^2-1} + \frac{\dd y^2}{1-y^2} \right) + (1-y^2) \dd{\phi}^2 \right]\nonumber \,,
\end{align}
where $t \in (-\infty, +\infty)$, $x \in (1, +\infty)$, $y \in [-1,1]$, and $\phi \in [0, 2\pi]$. In this metric, the location of the horizon and the singularity are at $x=1$ and $x=-1$, respectively. The ISCO (innermost stable circular orbit) for the usual, undistorted Schwarzschild space-time is at $x=5$. The relation to Schwarzschild coordinates is given by
\begin{align}\label{transf1}
x & =\frac{r}{M}-1,\\
 y & = \cos\theta. \label{transf2}
\end{align}
The distortion functions can be expressed in terms of Legendre polynomials of the first kind \cite{Chandrasekhar:579245,1997PhLA..230....7B}, 
\begin{align}
\hat{\psi} & = \sum_{n> 0} a_n R^n P_n, \label{say} \\
\hat{\gamma} & = \sum_{n>0} a_n \sum_{l=0}^{n-1} \left[ (-1)^{n-l+1} (x+y)-x+y \right] R^l P_l \nonumber \\
& \quad + \sum_{k,n=1} \frac{n k a_n a_k}{(n+k)} R^{n+k} [P_n P_k -P_{n-1} P_{k-1} ]\,, \label{gamma}
\end{align}
where 
\begin{align}
P_n:= P_n(xy/R)\,, \quad R=\sqrt{x^2+y^2-1}\,.
\end{align}
Where, $a_n\in \mathbb{R}$ are called multipole moments defining a distortion due to an external field. For $a_n=0$ we recover the Schwarzschild space-time, $a_1$ is the dipole moment, $a_2$ the quadrupole moment, and so on. The mass multiple moments constitute the deviations from the spherically symmetric shape of the central compact object \cite{1990ForPh..38..733Q}.
%Coordinate invariant of multipole moments were investigated by Geroch [7]also, with Beig and Simon [14] among others.

To fulfill the elementary flatness condition that is related to the requirement of having no singularities on the symmetry axis, we have to demand
\cite{1982JMP....23..680G}
\begin{equation}\label{equi}
\sum_{n>0} a_{2n-1} = 0 \,.
\end{equation}
Also, for an astronomical object the series \eqref{say} and \eqref{gamma} should converge, implying the existence of a real positive constant $k$ such that $|{a_n}|\leq{k}<\infty$ for all $n$.

In the following, we will restrict our consideration to a distortion up to the quadrupole moment $a_2$, assuming contributions of higher orders can be neglected. Moreover, for simplicity, we adapted the notation $q$ for the quadrupole moment, instead of $a_2$.

In this restriction to the quadrupole moment, one can think of some physical intuition from the Newtonian setup in the way that a positive quadrupole, modeling a distortion of the black hole due to a disk like structure, whereas a negative quadrupole models a mass distribution along the axis of symmetry. %by convention.

%........................................333.....................................................

\section{Thin accretion disk }\label{sec:disk}

In this section we introduce the structure of thin accretion disc and state the assumptions and equations in terms of $(t, r, \theta, \phi)$ coordinate as usual, for simplicity to be able to concentrate on the concept and for easy comparison with other papers. One could use the transformations \eqref{transf1} and \eqref{transf2} to rewrite them in distorted Schwarzschild coordinate easily.

The standard thin disk model (Shakura \& Sunyaev 1973 \cite{1973A&A....24..337S}, Novikov and Thorne 1973 \cite{1973blho.conf..343N}) can be applied to geometrically thin, optically thick, and cold accretion disks. One of the basic assumptions of this model is the viscosity mechanism, that generates heat which can be radiated away locally. 
In this model, the disk can radiate a considerable fraction of its rest mass energy. The mass accretion rate is assumed to be rather low, around $10\%$ of the Eddington mass rate ${M}_{\rm{Edd}}(=16L_{\rm{Edd}}c^2)$, where $L_{\rm{Edd}}$ is the Eddington luminosity. 

In the most of these models all physical quantities depend on the radial distance from the central object, and the vertical distance from the equatorial plane. Then the two-dimensional disk structure turns to two one-dimensional configurations: a radial quasi-Keplerian flow and a vertical hydrostatic structure. Also, in this model usually vertically integrated quantities is used.

For example, if $B$ is defined to be some quantity that is obtained by vertical integration of the quantity $b$, then we have
\begin{equation}\label{example}
{B(x)}:=\int^{+H(x)}_{-H(x)}b(x, z)\dd z, 
\end{equation}
where $H$ is half of the thickness of the disk.

%.............................................22-11........................................

\subsection{Thin disk approximations}\label{assumptions}

Before describing the basic equations in this model, meaning the conservation laws, equation of state, and radiation laws, we find it useful to collect and review the simplifying assumptions we use to solve the equations. These assumptions were introduced and discussed in the seminal papers by  Shakura and Sunyaev \cite{1973A&A....24..337S}, Novikov and Thorne \cite{1973blho.conf..343N}, and Page and Thorne \cite{1974ApJ...191..499P}.

\begin{itemize}
\item The disk is located in the equatorial plane, implying ${u}^{\theta}$ vanishes.
\item In the equatorial plane Keplerian circular orbits are assumed with a small radial drift velocity $u^r$, quasi-Keplerian, which is much smaller than the angular velocity. Near the central compact object $u^r$ is negative and gives rise to the mass accretion.
\item All radiation is vertical.
\item The disk is geometrically thin, $h=\frac{H}{r}\ll1$, where $H$ is half of the thickness of the disk. This assumption requires the disk to be cool and it implies $kT\ll \frac{GMm_p}{r}$, where $k$ is the Boltzmann constant, $T$ is the temperature of the disk, and $m_p$ is the mass of the proton.
\item The specific internal energy density, the radial pressure gradient, and therefore the advection are negligible.
\item The standard $\alpha$ viscosity prescription of the Shakura-Sunyaev model is assumed,
\begin{align}\label{VS}
S_{r \phi}=\alpha P,
\end{align} 
 where $S_{r \phi}$ is the only non-vanishing component of viscous (internal) stress energy tensor and $P$ is pressure. %And the subordinate shear tensor is $\sigma_{{\hat{r}\hat{\phi}}}=\sigma_{{\hat{\phi}\hat{r}}}$.  
 We explain more about this in sec.\ref{equations}.
\item %For the boundary condition at the 
For the inner edge of the disk, we use the "no-torque" boundary condition at the innermost stable circular orbit (ISCO), see e.g.~\cite{1973blho.conf..343N}. Although this seems to be a reasonable assumption in this case, there are some attempts to consider torque at the inner edge of the thin disks, see e.g.~\cite{2012MNRAS.420..684P}.
\end{itemize}

%............................... 22-22..................................

\subsection{Equations of the thin disk model}\label{equations}

Up to now, we adopted $c=1$ and $G=1$. In this section in addition we use $M=1$ for representation of the equations. 

There are three fundamental equations that govern the radial structure of thin disk model.

First, the particle number conservation
\begin{equation}\label{restmasscon}
(\rho u^{\mu})_{;\mu}=0 \,,
\end{equation}
where $u^{\mu}$ is the four-velocity of the fluid and $\rho$ is the rest mass density. The mass accretion rate is connected to this conservation law, meaning we expect the mass accretion rate to be constant, as otherwise we would see matter pile up at some certain region of the disk. 

Second, the radial momentum equation, a component of the relativistic Navier-Stokes equations, reads as
\begin{equation}\label{NSE}
h_{\mu \sigma}(T^{\sigma \nu})_{;\nu}=0 \,,
\end{equation}
where $T^{\sigma \nu}$ is the  stress-energy tensor, and $h^{\mu \nu} = u^{\mu} u^{\nu} + g^{\mu \nu}$ is the projection tensor giving the spatial metric normal to $u_{\mu}$. 

Finally, the energy conservation equation reads
\begin{equation}\label{energycon}
u_{\mu} T^{\mu \nu}{}_{;\nu}=0\,.
\end{equation}
The stress-energy tensor $T^{\sigma \nu}$ is given by 

\begin{align}
T^{\mu\nu}=hu^{\mu}u^{\nu}-Pg^{\mu\nu}+q^{\mu} u^{\nu}+q^{\nu} u^{\mu}+S^{\mu\nu},
\end{align}
where $h$ is enthalpy density, which is sum of internal energy  per unit proper volume and pressure over rest mass density, $P$ is the pressure, $u^{\mu}$ is four velocity of fluid, $q^{\nu}$ is transverse energy flux and $S^{\mu\nu}$ is the viscous stress energy tensor. In relativistic form, when we have no bulk viscosity, it is given by $S^{\mu\nu}=-2\lambda \sigma^{\mu\nu}$, where $\lambda$ is the dynamical viscosity and $\sigma^{\mu\nu}$ is the shear tensor.

In the used coordinates \eqref{transf1}-\eqref{transf2}, and according to the assumptions sec. \ref{assumptions}, the only non-vanishing component of the shear tensor is $\sigma_{r\phi}$. It is given in terms of the projection tensor $h^{\mu \nu}$ and the four-velocities as
\begin{align}\label{shearsigmahat}
\sigma_{r\phi}=\frac{1}{2}(u_{r;\beta}h^{\beta}_{\phi}+u_{\phi;\beta}h^{\beta}_{r})-\frac{1}{3}h_{r\phi}u^{\beta}_{;\beta}.
\end{align}
% In fact the two last equations are component of conservation of energy-momentum tensor parallel to four velocity and radial component of its projection with ${h}_{\mu \nu}$, respectively.

By imposing the assumptions of the thin disk model on the basic equations \eqref{restmasscon}-\eqref{energycon} together with the radiative energy transport law and the equations of state and opacity, we can derive the system of non-linear algebraic equations governing the thin disk model \cite{1973blho.conf..343N,2017MNRAS.468.4351C}, as follows:

The surface density $\Sigma$, is obtained by vertical integration of the density, see \eqref{example},

\begin{align}
\Sigma= 2\rho H, \label{sigma2}
\end{align}
where $H$ is disk height or half of the thickness of the disk.

In the context of accretion disks, the radial velocity of fluid which is responsible for accreting mass is obtained in terms of the mass accretion rate $\dot{M}$,
\begin{align}
u^r=-\frac{\dot{M}}{2\pi r \Sigma} \label{massrate},
\end{align}
One of the consequences of assumption of thinness of the disk, as mentioned earlier, is, there is no heat flow within the disk, except in the vertical direction. It means that we can only consider $q^z$ component. Therefore, the time-averaged flux of radiant energy $F$ (energy per unit proper area and proper time) flowing out of the upper and lower faces of the disk, is related to the  $z$-component of the energy flux, as $q^z(r,z) = F(r) \frac{z}{H(r)}$ [4, 37]. (Notice that here the notation $q^z$ is used for the usual energy flux in the stress-energy tensor, however, it's not related to the symbol $q$ that is taken for the quadrupole).
By using the fundamental equations \eqref{restmasscon},\eqref{NSE},\eqref{energycon} we get
\begin{align}
\frac{(\Omega {L}-E)^2}{\Omega_{,{r}}}\frac{{F{r}}}{{\dot{M}}}= \int_{{r}_0}^{r} \frac{(\Omega {L}-{E})}{4\pi}{L}_{,{r}}\dd{r}, \label{ene}
\end{align}
where $E=-u_t$ and $L=u_\phi$ are the energy and angular momentum per unit mass of geodesic circular motion in the equatorial plane, and $\Omega=\frac{u^{\phi}}{u^t}$ is the corresponding angular velocity.

For the vertically integrated viscous stress ${W}$, one should integrate the viscous stress energy component $S_{r\phi}$, sec.~\ref{assumptions} equation \eqref{VS}, vertically through the disk by using the relation \eqref{example}, which we get
\begin{equation}\label{w}
{W}= 2{\alpha}P {H},
\end{equation}
where the pressure $P$, in general, is the sum of gas pressure (from nuclei) and radiation pressure,

\begin{equation}\label{P}
P=\frac{{\rho kT}}{{m_p}}+\frac{{a}}{{3}}{T}^4,
\end{equation}
where ${m_p}$ is the rest mass of the proton, ${k}$ is Boltzmann's constant, $a$ is the radiation density constant, and $T$ is the temperature. In fact we ignore the mass difference between neutron and proton in the first term for simplicity. In practice, the pressure balance equation in the vertical direction can be obtained by 

\begin{equation}\label{VP}
\frac{P}{{\rho}}=\frac{1}{2}\frac{{(HL)^2}}{{r^4}}\,,
\end{equation}
which is based on the work of Abramowicz et al. \cite{1997ApJ...479..179A}.

The vertical configuration is mainly governed by the pressure, also by energy transportation law, which is
\begin{equation}\label{OD}
4\sigma {T}^4={\Sigma} {F}{\kappa},
\end{equation}
where $\sigma$ is Stefan-Boltzmann’s constant and $\kappa$ is Rosseland-mean opacity,
\begin{align}
\kappa=0.40 + 0.64\times10^{23} (\frac{\rho}{g \hspace{0.1cm}cm^{-3}})(\frac{T}{K})^{-\frac{7}{2}} cm^2 \hspace{0.1cm}g^{-1}.
\end{align}
Here the first term is electron scattering opacity and the second one is free-free absorption opacity.

The final equation, states the relation between the flux $F$, and the viscous stress tensor $W$,
\begin{align}
{F} = -{\sigma_{\hat{r}\hat{\phi}}} {W}.\label{navi}
\end{align}

Where the \textit{hat} symbols refer to the local rest frame of the fluid. To obtain the solutions for the this model one needs to solve equations \eqref{sigma2}-\eqref{navi} which are responsible for the dynamics of the thin accretion disk.

%................................................444................................................

\section{Thin accretion disk around the distorted Schwarzschild black hole}\label{sec:eq}

\subsection{Construction of the disk}

From now on, as we mentioned before, $q$ refers to the quadrupole $a_2$ for simplicity. As the first step, we rewrite the metric for the distorted black hole solution \eqref{Dmetric} in the equatorial plane, meaning $\theta =\frac{\pi}{2}$ or equivalently ${y}=0$ in the coordinates \eqref{transf1}. In the equatorial plane the distortion functions \eqref{say}, \eqref{gamma} up to the quadrupole moment are,
\begin{align}\label{gammapsi1}
	\hat{\psi} & = -\frac{q}{2}({x}^2-1)\,, \\
\label{gammapsi2}
	\hat{\gamma} & = -2q{x}+\frac{q^2}{4}({x}^2-1)^2\,.
\end{align}
Therefore, the metric \eqref{Dmetric} turns to,
\begin{align}
	\dd{s}^2= &-\left( \frac{{x}-1}{{x}+1} \right) e^{2\hat{\psi}}\dd{t}^2\\
	&+ {M}^2 ({x}+1)^2 e^{-2\hat{\psi}}\left[ e^{2\hat{\gamma}} \left( \frac{\dd{x}^2}{{x}^2-1}+\dd{y}^2 \right) + \dd{\phi}^2 \right]\nonumber\,.
\end{align}
By the substitution $q=0$ we retrieve the usual Schwarzschild metric. 

As the second step, one can derive the various physical quantities appearing in the equations of the thin disk model described earlier in sec. \ref{equations}. 

For circular geodesic motion, the total energy and angular momentum per unit mass as well as the angular velocity in this space-time are
\begin{align}
    E & = \frac{(x-1)e^{-\frac{q}{2}(x^2+3)}}{\sqrt{x+1}} \sqrt{\frac{qx^2+qx+1}{2x^3q+x(1-2q)-2}}\,, \label{Energy}\\
    L & =M(x+1)e^{\frac{q}{2}(x^2+3)}\sqrt{\frac{-qx^3+qx+1}{2x^3q+x(1-2q)-2}}\,,\label{angmom}\\
    \Omega & = \frac{e^{-q(x^2+3)}}{M\sqrt{(x+1)^3}} \sqrt{ \frac{-qx^3+qx+1}{qx^2+qx+1} }\,. \label{Omega}
\end{align}
By its definition the distorted metric is only valid locally, in some neighborhood of the horizon. We can quantitatively restrict this region of validity by observing the behaviour of physical quantities $E, L$ and $ \Omega$ appearing in the basic equations. Meaning, we consider these quantities, only in the region that they have real values. This gives us some restrictions on the range of the ${x}$ coordinate for a given quadrupole moment.

In addition, for a vanishing quadrupole $q$ the angular velocity $\Omega$ is a monotonically decreasing function of the radius. However, in this distorted space-time, $\Omega$ may have an extremum. As the black hole is surrounded by
a mass distribution, it is not a big surprise that after some distance the
behaviour of $\Omega$ manifests this existence of an external matter distribution. At this extremum, the influence of the surrounding matter becomes strong, indicating that it is close enough that the local solution is no longer valid. Therefore, we take the appearance of this extremum as another restriction to the
validity of the local black hole solution.

In the case of $q<0$ this extremum in $\Omega$ appears within the valid range of radial coordinates where $E$, $L$ and $\Omega$ are real. However, for $q>0$ this extremum appears outside this valid range that one can obtain from the above complexity analysis, and therefore plays no role in this case.

Let us first discuss positive quadrupole moments $q>0$. In this case, the polynomial $(qx^2+qx+1)$ in eq.~\eqref{Omega} is always positive for $x>0$ therefore, we have just the following condition
\begin{align}\label{positiveq}
  -qx^3+qx+1 > 0 \,.
\end{align}
As this polynomial has always exactly one real positive zero by Descartes' rule, this condition gives us an upper bound on $x$, that is always larger than the horizon, meaning $x=1$. The second condition $2x^3q+x(1-2q)-2>0$ from the equations \eqref{Energy} and \eqref{angmom} gives us nothing new, because it gives only a lower bound on $x$, that is between $x=1$ and $x=2$ and therefore smaller than the ISCO of undistorted Schwarzschild at $x=5$, see also the discussion of the inner edge below.

In the case of a negative quadrupole $q<0$, for deriving the first restriction one need to analyze $L$ and $\Omega$. However, as we mentioned before, by analysing the behaviour of $\frac{\partial{\Omega}}{\partial{x}}$ one can show that for $0>q\gtrsim -0.021$, within the valid region where the physical quantities $L$, $E$ and $\Omega$ are real, $\Omega$ always has an minimum. Here $q\approx -0.021$ is a double zero of $2x^3q+x(1-2q)-2>0$, and also seems to coincide with the minimum valid quadrupole for the existence of an ISCO, see \cite{2016PhRvD..93f4019S} and sec. \ref{ueffsec}. There is no compact analytical form for the minimum of $\Omega$, and we find it numerically for a given $q$. So the place of the extremum determines the valid range for the radial coordinate $x$ in the case of a negative quadrupole.

In summary, in the case of a positive quadrupole moment, we find a maximum value for the radial coordinate $x$ from the complexity analysis of eqs.~\eqref{Energy}-\eqref{Omega}. For a negative quadrupole moment, we find the maximum value of $x$ from the location of the minimum of angular velocity \eqref{Omega} for a given $q$.% and we don't need to take into account analysis of eqs.\eqref{fromomega} and \eqref{frome}.

We should mention that as the absolute value of the quadrupole gets smaller and smaller, we get a wider range for $x$, as expected from the Schwarzschild limit.

Another quantity that is calculated from the metric is $\sigma_{\hat{x}\hat{\phi}}$ which is the shear rate in the local rest frame. Aside from the fact that we are dealing with local radiation in the standard thin disks, this frame is suitable to do small modifications of an inviscid flow. The relation between $\sigma_{\hat{\nu}\hat{\mu}}$ and $\sigma_{\nu\mu}$ is given by

\begin{align}
\sigma_{\hat{\nu}\hat{\mu}}=e^{\nu}_{\hspace{0.1cm}\hat{\nu}}e^{\mu}_{\hspace{0.1cm}\hat{\mu}}\sigma_{\nu\mu},
\end{align}
where the basis $e^{\nu}_{\hspace{0.1cm}\hat{\mu}}$ contains orthonormal vectors. This basis is scaled by the coefficient in the original metric to obtain a unit vector. %$e^{\nu}_{\hat{t}}$ is parallel to $u^{\nu}$ meaning  $e^{\nu}_{\hat{a}}u_{\nu}=\delta^{\hat{t}}_{\hat{a}}$, and the basis vectors are orthonormal.
As we had before \eqref{shearsigmahat}, $\sigma_{x\phi}$ is given by

\begin{align}
\sigma_{x\phi}=\frac{1}{2}(u_{x;\beta}h^{\beta}_{\hspace{0.1cm}\phi}+u_{\phi;\beta}h^{\beta}_{\hspace{0.1cm}x})-\frac{1}{3}h_{x\phi}u^{\beta}_{;\beta}.
\end{align}
%As it sounds, it is a long calculation in which one should calculate many of the pieces of
The long calculation of $\sigma_{x\phi}$, contains terms from the projection tensor $h_{\nu\mu}$ \eqref{NSE}, the covariant derivative of the four velocity, and the Christoffel symbols,

\begin{align}
\Gamma^t_{tx}&=\frac{1}{x^2-1}+\hat{\psi}_{,x},\\
\Gamma^{\phi}_{x\phi}&=\frac{1}{x+1}-\hat{\psi}_{,x},\\
\Gamma^x_{xx}&=\frac{-1}{x^2-1}(\hat{\gamma}_{,x}-\hat{\psi}_{,x}),\\
\Gamma^x_{tt}&=\frac{(x-1)(1+\hat{\psi}_{,x}(x^2-1))}{(x+1)^3M^2}e^{4\hat{\psi}-2\hat{\gamma}},\\
\Gamma^y_{yx}&=\frac{1}{x+1}+\hat{\gamma}_{,x}-\hat{\psi}_{,x}, 
\end{align}
where $\hat{\gamma}$ and $\hat{\psi}$ are given by eqs.~\eqref{gammapsi1} and \eqref{gammapsi2}.

%...............................................
\subsection{Inner edge of the thin disk}\label{ueffsec}

The inner edge of the standard thin disk model is assumed to be at the Innermost Stable Circular Orbit (ISCO), also called marginally stable orbit. So, it is natural to analyse the location of the ISCO in the distorted Schwarzschild space-time in the next step. Because of reflection symmetry we have the following condition for existence of geodesics in the equatorial plane 
% \old{In the analysis of geodesic equations imposing $y=0$ that gives us geodesics confined to the equatorial plane, and ask also first and second derivative of $y$ to be vanished along this geodesics, it turns out that there exist geodesics confined to the equatorial plane if }
\begin{align}
  a_{2n-1}=0 \text{ for } n>0\, .
\end{align}
We see this condition is actually stronger than the previous condition \eqref{equi}; however, in the study of quadrupole, it gives us nothing new. Also, from the analysis of the effective potential for this distorted Schwarzschild black hole,

\begin{align}
    V_{\rm Eff}= \left( \frac{x-1}{x+1} \right) e^{2\hat{\psi}} \left( 1+\frac{L^2e^{2\hat{\psi}}}{M^2(x+1)^2} \right)\,,
\end{align}
it was shown that the ISCO\old{and bound orbits} exists only for ${q}\in ({q}_{\rm min}, {q}_{ \rm max}]$, where ${q}_{\rm min}\approx -0.0210$ and ${q}_{\rm max}\approx  2.7086\times 10^{-4}$ \cite{2016PhRvD..93f4019S}. In the case of a negative quadrupole $q<0$ the ISCO is closer to the horizon, as compared to the undistorted Schwarzschild case at $x=5$, with the smallest ISCO given by $x \approx 2.8794$. This is about $37\%$ closer to the horizon than in the Schwarzschild case \cite{2016PhRvD..93f4019S}. For a positive quadrupole, the ISCO is located at larger radial coordinates from the horizon with respect to the ISCO in the undistorted Schwarzschild case. Considering the largest ISCO, it is at $x \approx 6.5018$, about $25\%$ farther away.

Therefore, from an ISCO analysis we got a restriction on choosing the  quadrupole moment in the model. 

At this step before solving the system of equations \eqref{sigma2}-\eqref{navi}, one needs to rewrite all equations and assumptions in the $(t, x, y, \phi)$ coordinates which is introduced earlier.
%...................................................
\section{Results from the thin disk model}\label{sec:results}
By rewriting the whole system of equations in sec. \ref{equations}, in the distorted space-time and manipulating them by the mentioned assumption and the restrictions, we obtain the system of algebraic equations. Our results are described and plotted in the following. Also it is interesting to compare these results with the usual undistorted Schwarzschild space-time. We should mention that as the distorted solutions are only valid locally, in the vicinity of the horizon, we only consider the inner part of the disk in this space-time. Up to which point we can exactly consider the solution depends on the choice of the  quadrupole, as explained in section \ref{sec:eq}. Moreover we assume electron scattering opacity and radiation pressure dominant \cite{Abramowicz2013}. Our results were produced using the following dimensionless variables for black hole mass and accretion rate:

\begin{figure}
    \includegraphics[width=0.45\textwidth]{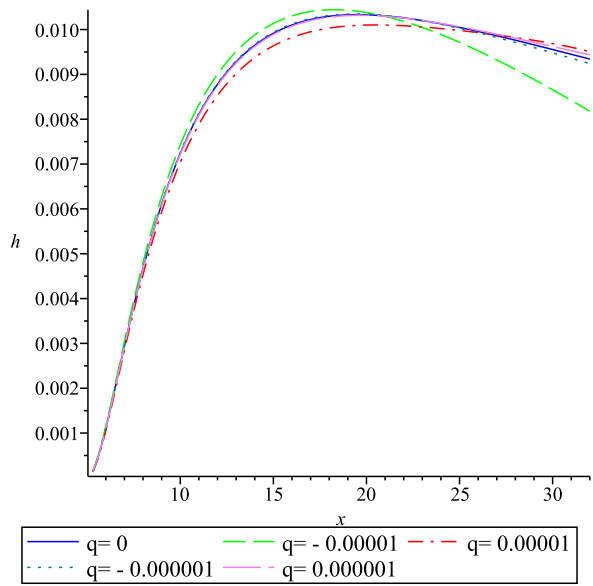}
    \caption{\label{hplot}Height scale $h$ of the disk for $q\neq0$ and $q=0$.}
\end{figure}

\begin{figure}
        \includegraphics[width=0.45\textwidth]{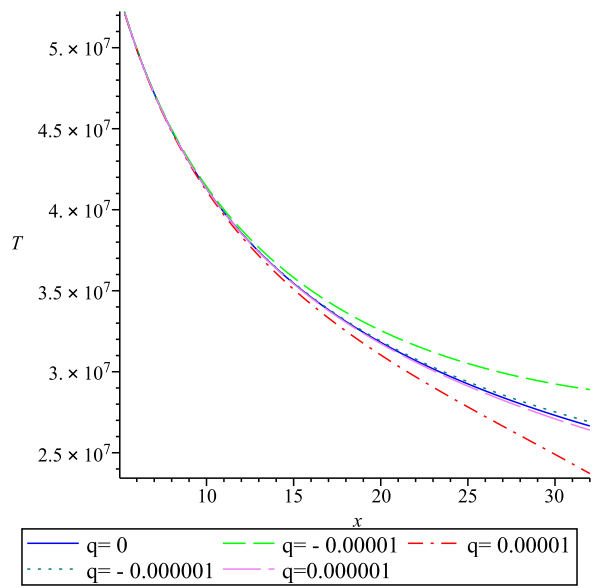}
    \caption{\label{tplot}Temperature $T$ for $q\neq0$ and $q=0$ in unit $\rm K$.}
\end{figure}

\begin{figure}
    \includegraphics[width=0.45\textwidth]{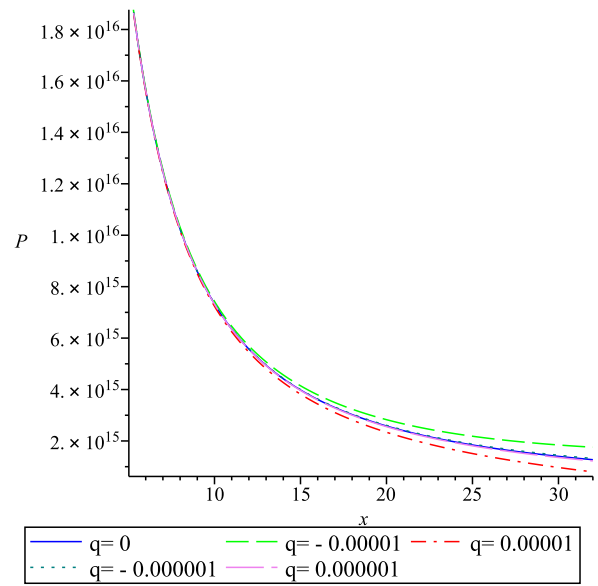}
    \caption{\label{pplot}Pressure $P$ for $q\neq0$ and $q=0$ in the $\rm \frac{dyn}{cm^2}$ unit.}
\end{figure}

\begin{align}
m&=\frac{M}{M_{\odot}}\,,\\ \dot{m}&=\frac{\dot{M}c^2}{L_{\rm{Edd}}}\,,
\end{align}
where $L_{\rm{Edd}}$ is Eddington luminosity, $L_{\rm{Edd}}\equiv1.2 \times 10^{38}\frac{M}{M_{\odot}} \hspace{0.2cm} {\rm erg \hspace{0.1cm} s^{-1}}.$
Also, for the numerical solutions produced in the following figures, the physical constants and parameters that were used are as follows, 
\begin{align}\label{p}
G&=6.67 \times 10^{-8} {\rm cm^3 g^{-1} s^{-2}},\\
c&= 3 \times10^{10} {\rm cm \hspace{0.1cm} s^{-1}}, \\
k &=1.38 \times 10^{-16} {\rm erg \hspace{0.1cm} K^{-1}}, \\
m_{p}&=1.672 \times 10^{-24} {\rm g},\\
\kappa_{es}&= 0.4 \hspace{0.1cm} {\rm cm^2 g^{-1}}.
\end{align}
and
\begin{align}\label{param}
M&={M_{\odot}}(\sim 1.99 \times 10^{33} \rm g),\\
\alpha&=0.1,\\
\kappa&=\kappa_{\rm es},\\
\dot{M}&=M_{\rm Edd}(=16L_{\rm Edd}c^2).\label{parame}
\end{align}
Aside from $h=\frac{H}{r}$, which is a dimensionless height scale, the results are represented in cgs units.

 With the parameters specified in \eqref{p}-\eqref{parame}, we solved the eight equations \eqref{sigma2}-\eqref{OD} analytically, and \eqref{navi} numerically, for the physical quantities $h$, $T$, $P$, $W$, $F$, $\Sigma$, $u^x$ and $\rho$. Below we show the results and discuss vertically integrated quantities. As it was mentioned in the assumptions, it is important that in this model, $u^x$ becomes negative, since it's responsible for accreting matter. Also, the information of $\rho$ is contained in surface density $\Sigma$, which is vertically integrated over it eq.\eqref{p}, and we discussed it below.

In figure \ref{hplot}, the height scale $h$ of the disk over $x$ is plotted. Obviously, smaller quadrupoles are closer to the Schwarzschild case, and as $|q|$ gets bigger the deviations increase. Close to the ISCO almost no difference between $q=0$ and any value $q \neq 0$ can be seen. But as the distance from the ISCO is increasing, we see strong deviation from the undistorted Schwarzschild case. In the latter the disk gets thicker and after reaching some maximum, very slowly gets thinner. However, we see this changing of the slope is less manifest for positive quadrupoles while it is stronger in the case of negative quadrupole. Moreover, for a negative quadrupole the radial position of the maximum height is shifted to smaller values, while the height of the maximum increases. The opposite can be observed for a positive quadrupole, the maximum is shifted to the right and its height decreases. 
However, as we saw in subsec. \ref{ueffsec}, in the case of negative quadrupole moment, ISCO is closer to horizon and we expect stronger deviation from the undistorted Schwarschild.

\begin{figure}
        \includegraphics[width=0.45\textwidth]{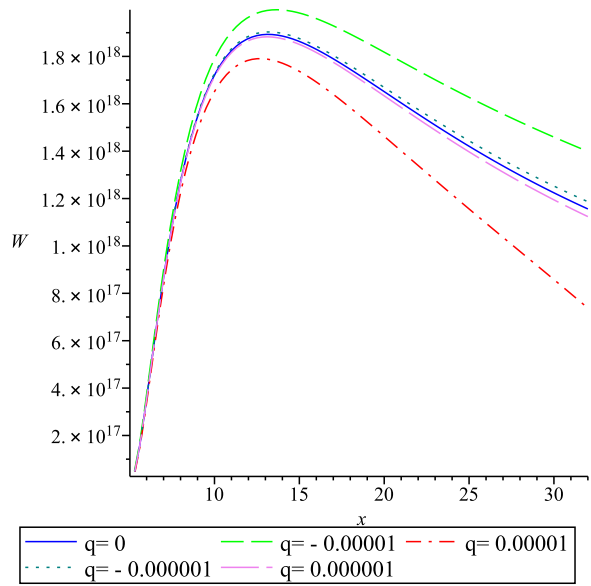}
     \caption{\label{wplot}Vertically averaged stress tensor $W$ for $q\neq0$ and $q=0$ in the $\rm \frac{dyn}{cm}$ unit.}
\end{figure}
\begin{figure}
    \includegraphics[width=0.45\textwidth]{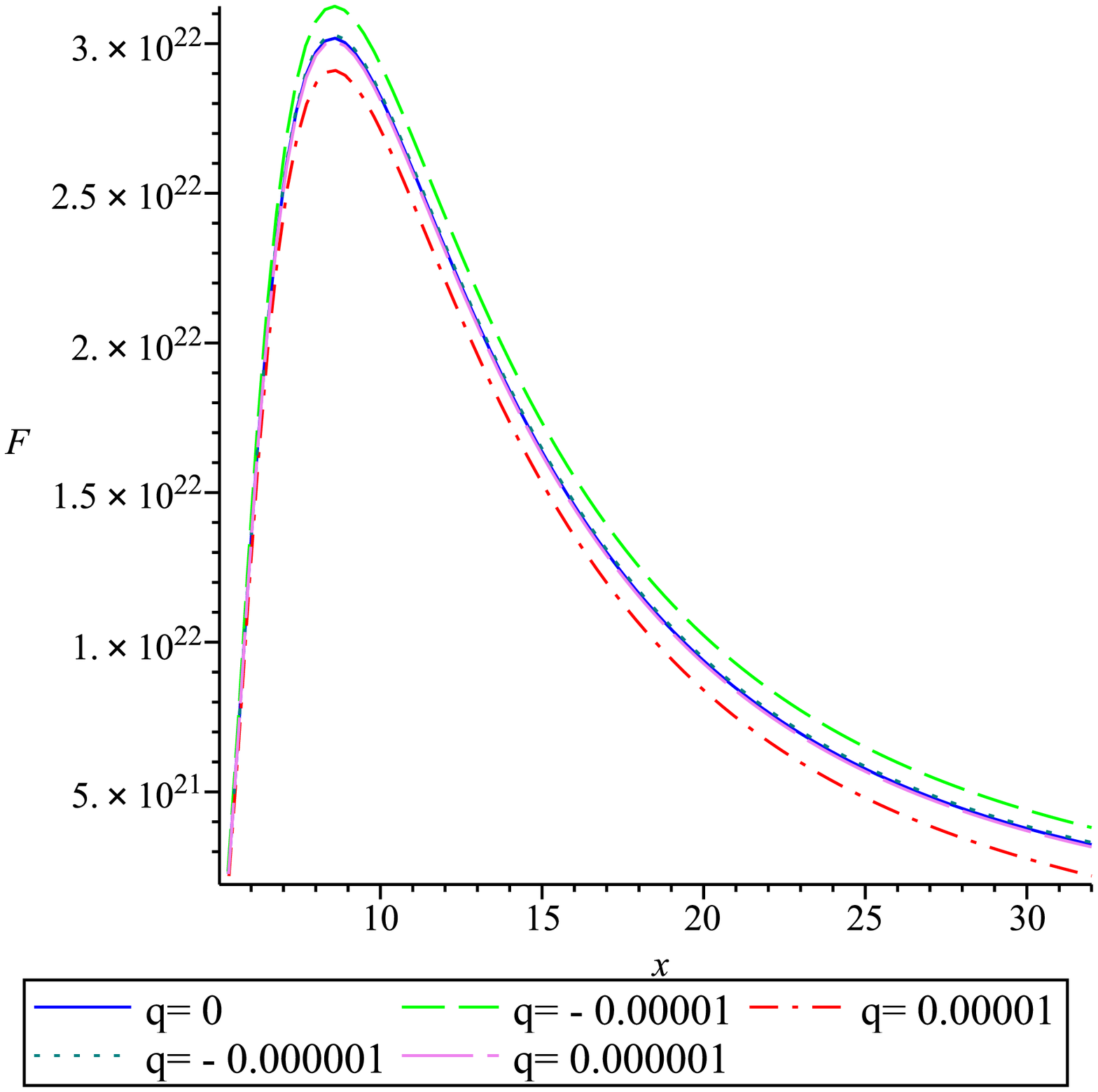}
    \caption{\label{fplot}Radiation flux $F$ for $q\neq0$ and $q=0$ in the $\rm \frac{erg}{cm^2s}$ unit.}
\end{figure}

The result shown in figure \ref{tplot} corresponds to the temperature $T$ over radius $x$. As we see, at small radii all plots for different values of $q$ are almost in agreement, and as $x$ is increasing we see some deviations. In the case of a positive quadrupole moment we see a more rapid descend at larger $x$ compared to the undistorted Schwarzschild, and for the negative quadrupole moment there is a slower descend with respect to undistorted Schwarzschild case. %Again, that is not a surprise because in the case of negative $q$, inner edge of the disk is closer to the horizon and we see stronger effects in inner part. \textcolor{red}{Can we really say that? The difference in the ISCO is only marginal, not even visible in the plot. Compared to that, the differences in $T$ are huge...Also, it would mean that the deviations for $q>0$ are weaker, but this is not the case for $T$, $F$ and $W$!?} 
The changes in temperature $T$ due to a non-vanishing quadrupole are similar to the results for the pressure $P$, shown in figure \ref{pplot}.
Again, in the case of a positive quadrupole moment we see a more rapid descend at larger $x$ and for the negative quadrupole moment there is a slower descend, with respect to the undistorted Schwarzschild case.
\begin{figure}
        \includegraphics[width=0.45\textwidth]{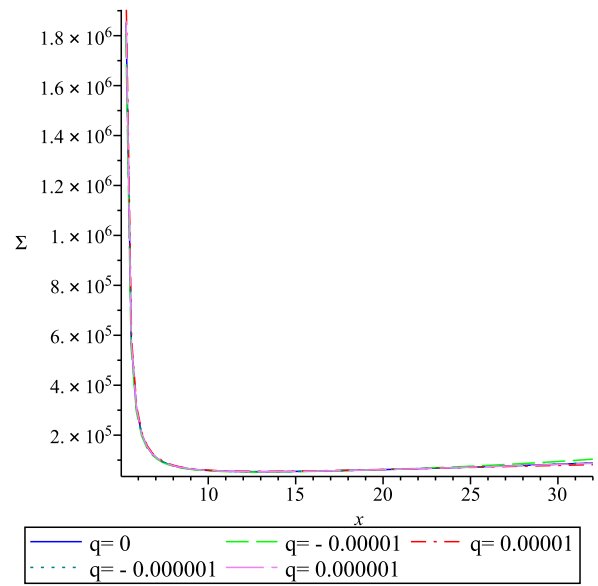}
        \caption{\label{splot}Surface density $\Sigma$ for $q\neq0$ and $q=0$ in the $\rm \frac{g}{cm^2}$ unit.}
\end{figure}
\begin{figure}
    \includegraphics[width=0.45\textwidth]{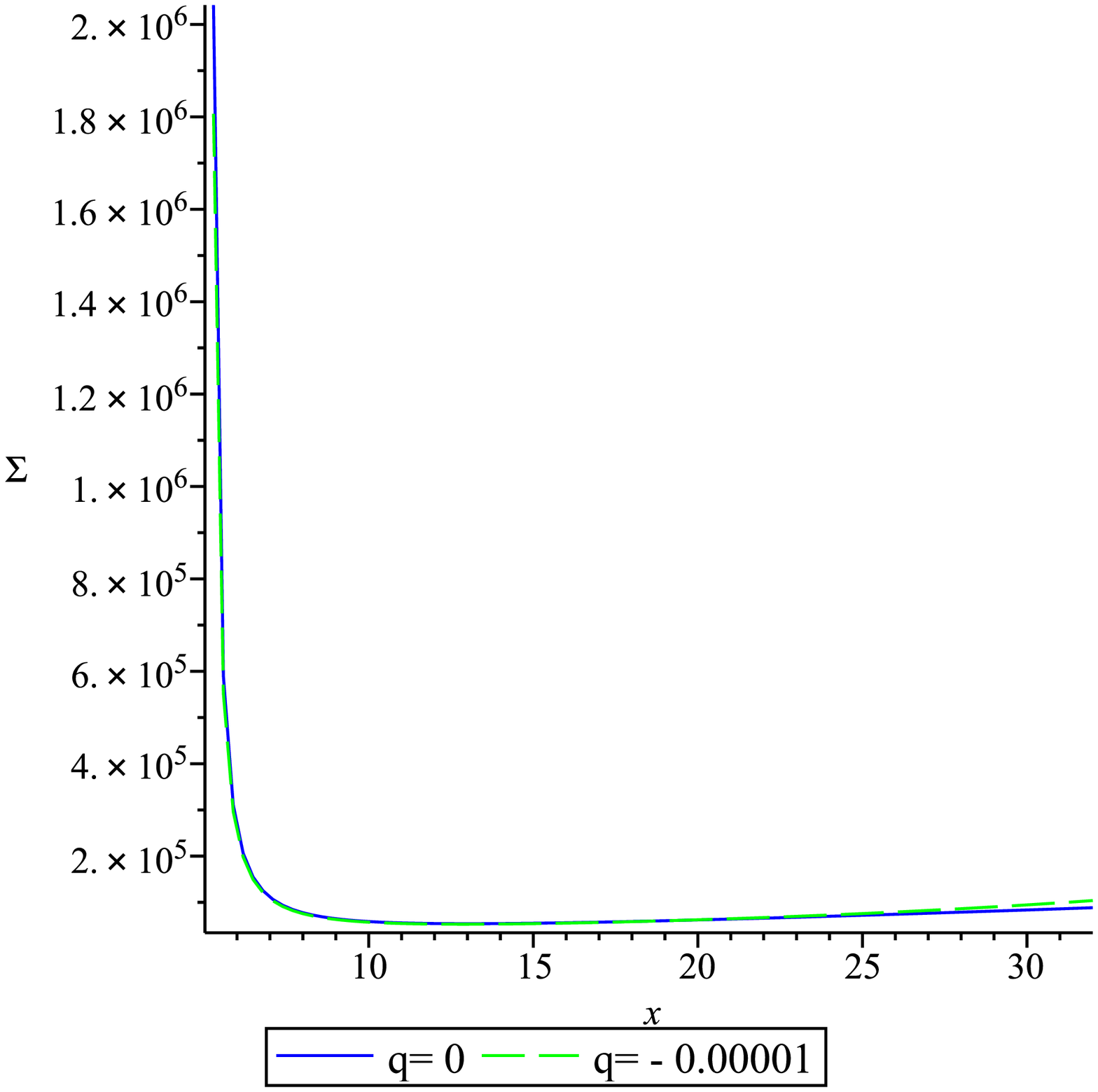}
     \includegraphics[width=0.45\textwidth]{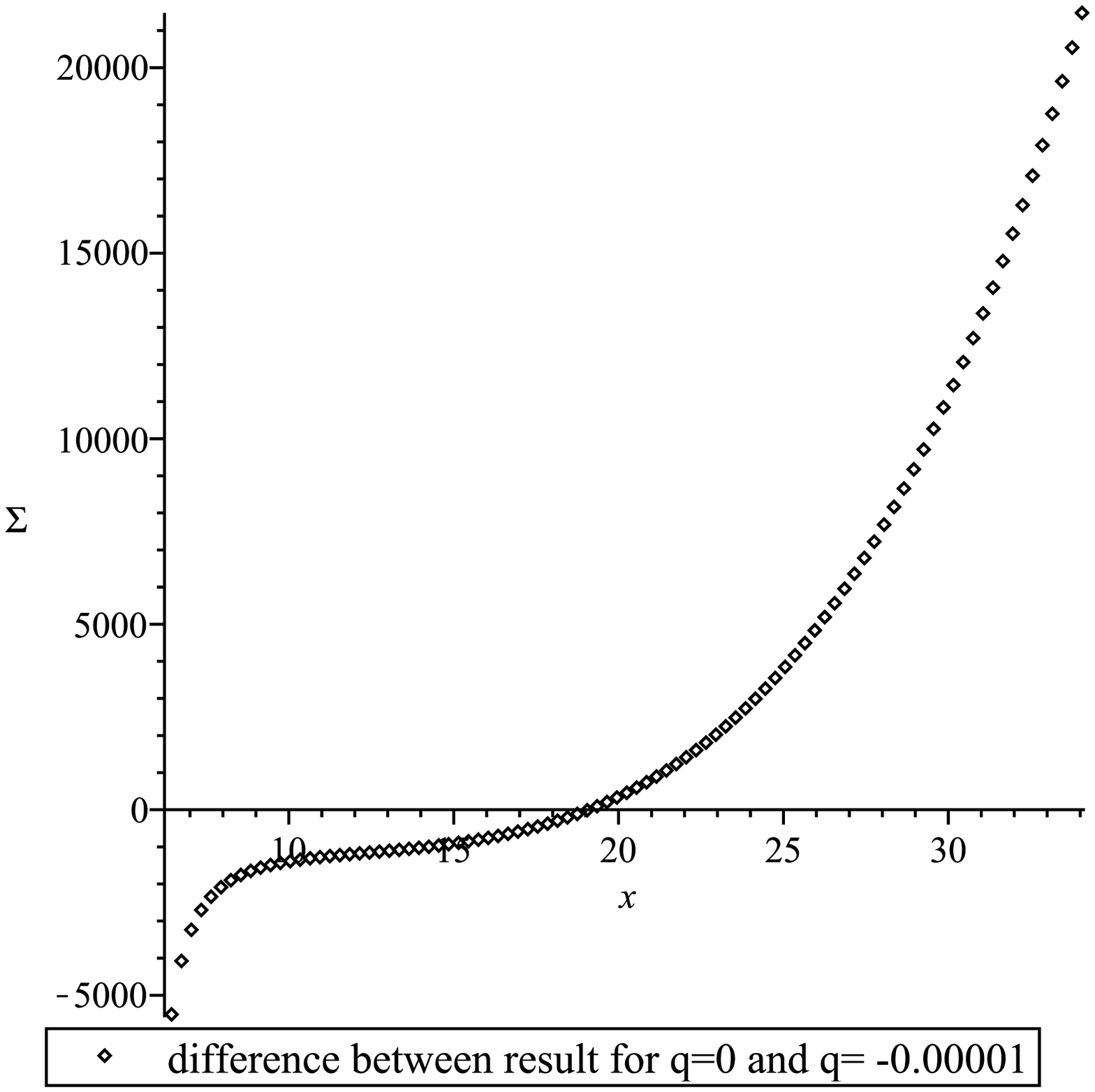}
    \caption{\label{sigmadifferenc01}Differences in surface density in the case of negative quadrupole moments and undistorted Schwarzschild.}
\end{figure}
\begin{figure}
        \includegraphics[width=0.45\textwidth]{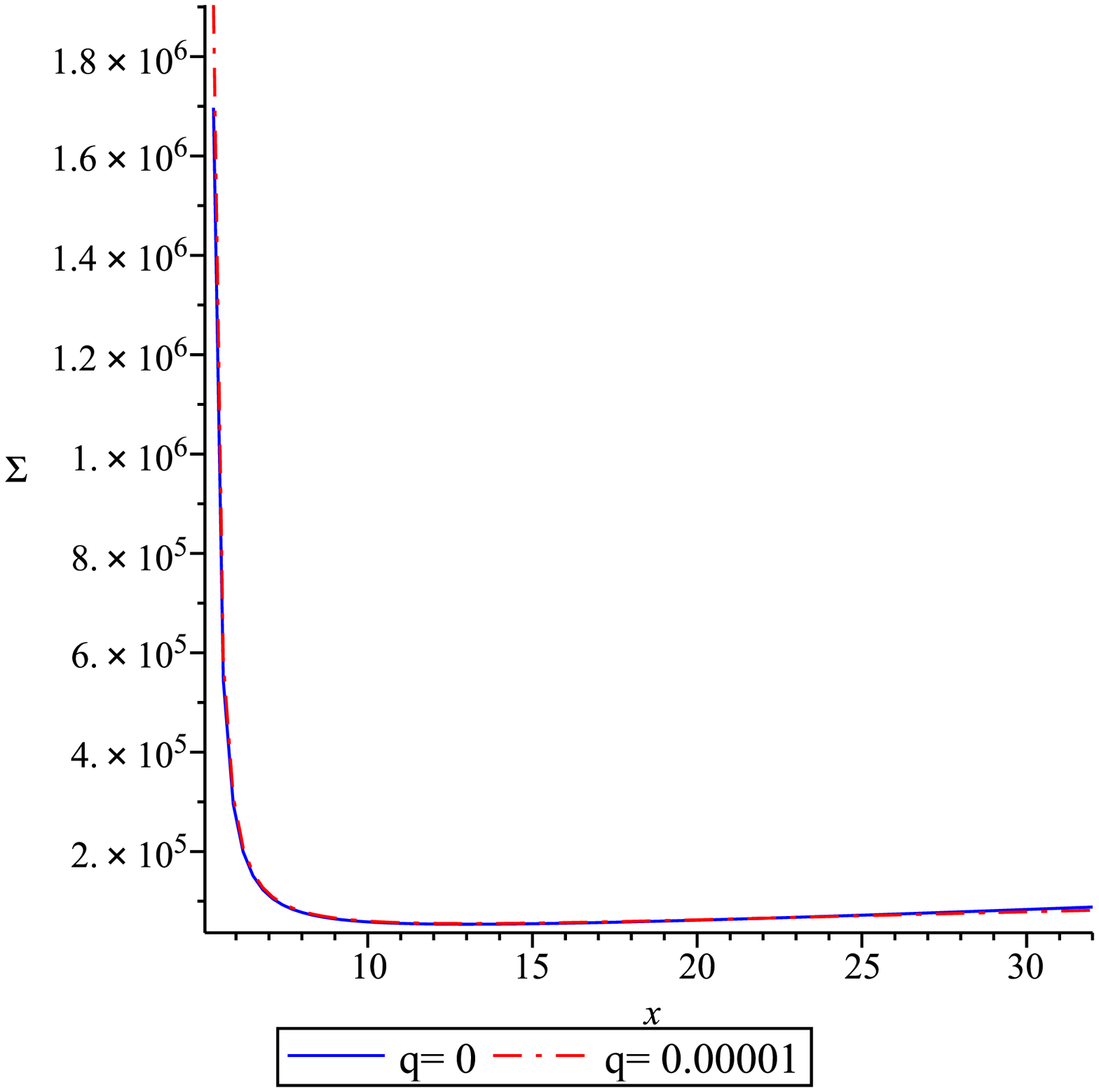}
         \includegraphics[width=0.45\textwidth]{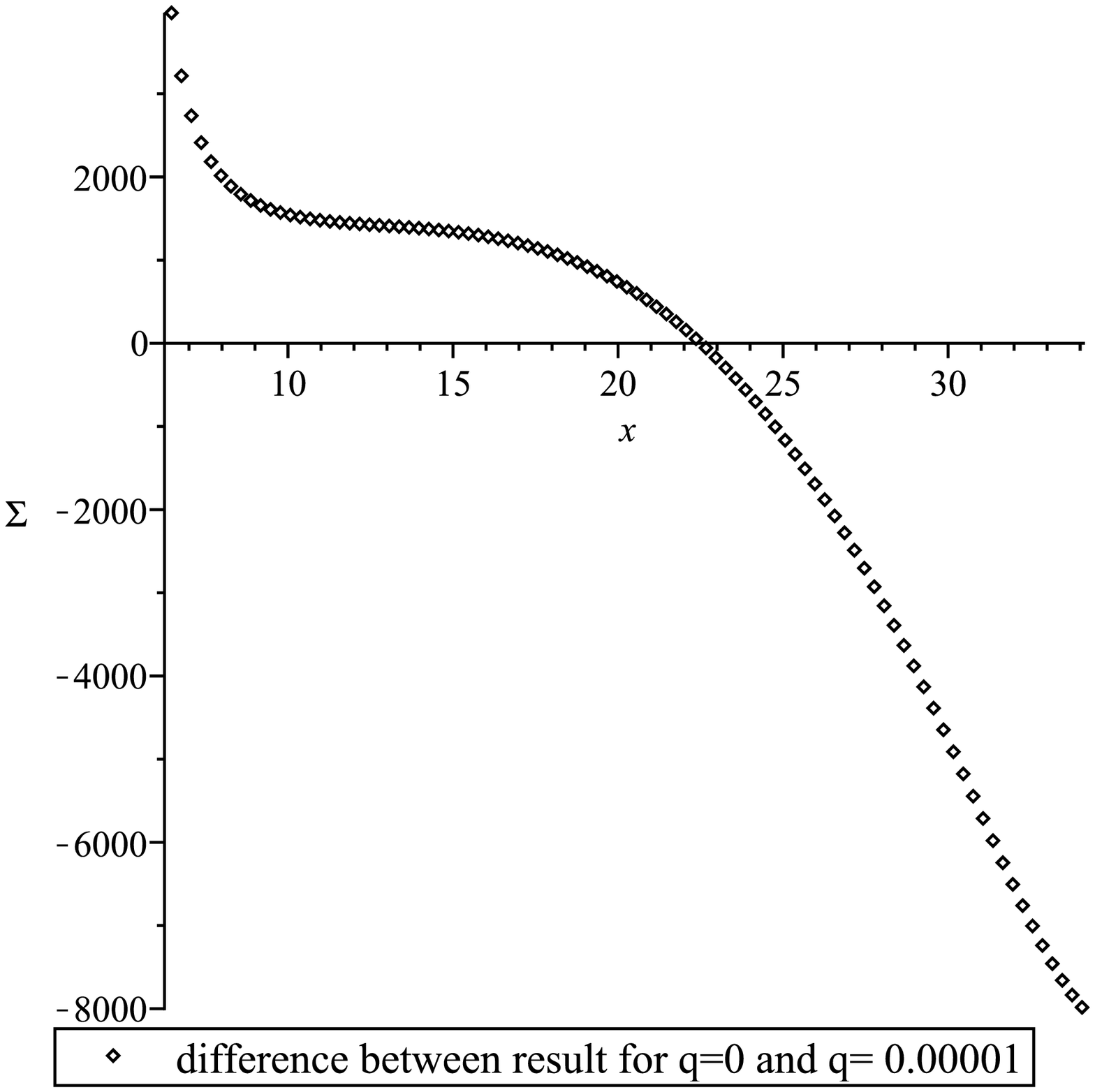}
    \caption{\label{sigmadifferenc02}Differences in surface density in the case of positive quadrupole moments and undistorted Schwarzschild.}
\end{figure}

In figure \ref{wplot} the vertically integrated viscous stress tensor $W$ is plotted, which is responsible for removing angular momentum from the disk. As before, we see at the small radii all plots for different values of $q$ are almost in agreement and as $x$ is increasing we see some deviations. The maxima of the plots are clearly shifted, in the opposite way than for the height scale $h$. For negative $q$ the maxima appear at a larger radial coordinate, and for $q>0$ at a smaller radial coordinate than in the undistorted Schwarzschild case $q=0$. However, in a similar way as for $h$, in the case of negative quadrupoles the maxima are higher than undistorted Schwarzschild, and lower for positive $q$. The overall shapes of the plots are all similar.

The radiation flux $F$, is plotted in figure \ref{fplot}. We see that the maxima of the plots are all nearly at the same position. Also, in the case of negative quadrupoles, $F$ is in general larger than in the undistorted Schwarzschild case, and in general is smaller in the case of positive quadrupoles.

Figure \ref{splot} shows the surface density $\Sigma$ over the radial coordinate $x$. As we see at very small radius, the plots for the undistorted Schwarzschild case $(q=0)$ and the distorted Schwarzschild case $(q\neq0)$  are almost in good agreement. However, as $x$ increases we see some deviations. The plots for positive quadrupoles are decreasing very slowly and the plots for negative quadrupoles are increasing very slowly, both compared to the undistorted Schwarzschild space-time. However, to see the differences better we discuss these two cases separately in the following figures \ref{sigmadifferenc01} and \ref{sigmadifferenc02}.

The case of a negative quadrupole is shown in figure \ref{sigmadifferenc01}. Here we plotted the difference $\Sigma(x,q<0) - \Sigma(x,q=0)$ of the surface density for $q=-10^{-5}$ with respect to the undistorted Schwarzschild space-time with $q=0$. We see that at radii close to the ISCO the difference is negative and then monotonically increases for larger radii. At large radii we therefore find the familiar behaviour that the values of $\Sigma$ are larger for negative quadrupole moment as compared to the Schwarzschild case. 

% In figure \ref{sigmadifferenc01}, surface density $\Sigma$, for a negative quadrupole moment $q=-0.00001$  with  respect  to  undistorted Schwarzschild  is  plotted.   Also,  the  difference  between $q= 0$ and $q=-0.00001$.  As we see at the small radii they have very small divergence; however, as $x$ increasing we see some larger deviations,  in which in the case of negative quadrupole the values of $\Sigma$ for a given $x$ is higher.

For a positive quadrupole moment we show the difference $\Sigma(x,q>0) - \Sigma(x,q=0)$ in figure \ref{sigmadifferenc02}. With our choice $q=10^{-5}$ the differences are generally smaller as for a negative quadrupole with the same absolute value, shown in figure \ref{sigmadifferenc01}. Apart from that, the behaviour is opposite to the case of negative quadrupole shown in figure \ref{sigmadifferenc01}. 
% In figure \ref{sigmadifferenc02}, surface density $\Sigma$, for a positive quadrupole moment $q=0.00001$ with respect to undistorted Schwarzschild is plotted. Also, the difference between $q=0$ and $q=0.00001$ is plotted. As we see at the small radii, again they have much smaller deviation; however, as $x$ is increasing we see stronger deviation, in which in the case of positive quadrupole the values of $\Sigma$ for a given $x$ is lower.

\section{Summary and Discussion}\label{sec:discuss}
%In this paper we have shown that, the formation of thin disk around distorted Schwarzschild black hole is restricted and has significant differences with thin disk around undistorted Schwarzschild black hole itself.
In this paper we constructed and discussed the standard relativistic thin disk model around a distorted Schwarzschild black hole. After reviewing the distorted space-time as well as the assumptions and equations governing the thin disk, we proceeded to analyse the structure of the equations for the case of a distorted black hole. As this solution is only locally valid in the vicinity of the horizon, we derived some physical conditions for the region of validity. Moreover, from the condition of existence of an ISCO, see \cite{2016PhRvD..93f4019S}, we found a valid range for the quadrupole parameter $q$. For different choices of $q$ from this range, both positive and negative, we derived the physical characteristics of the thin disk solution around the distorted Schwarzschild space-time and presented them in several figures. 

For producing all the figures in the paper, we fixed the parameters of the model, apart from the quadrupole $q$, to the values specified in eqs.~\eqref{param}-\eqref{parame}. For every choice of $q$ we of course obtain different results. We showed the surface density, the height scale of the disk, the temperature, the radiation flux, the pressure, and the viscous stress. For the surface density, we in addition presented the differences between the distorted and the undistorted Schwarzschild for two choices, negative and positive quadrupole moment, to see the deviation more clearly. In general, for all discussed examples, at small radii close to the ISCO a non-vanishing quadrupole, has a rather small impact on the results. The deviations between the Schwarzschild case $q=0$ and different choices of the quadrupole become more obvious for larger radii. For the monotonically decreasing quantities, temperature $T$ and pressure $P$, we commonly find that a negative quadrupole causes a smaller decrease compared to $q=0$, whereas a positive quadrupole gives rise to a stronger decrease. For the dimensionless height scale $h=H/r$, we find a shift of the radius of maximum height $h$ and a modified behaviour of the slope for larger radii as compared to the Schwarzschild case. Interestingly, for the viscous stress $W$, the position of the maximum is shifted for negative and positive $q$ in the opposite way as for $h$. Apart from that, for both $W$ and the radiation flux $F$, we find the already familiar behaviour that the results for a negative quadrupole lie generally above the Schwarzschild case, and below for positive $q$.  Lastly, the surface density $\Sigma$ is in some sense special. Generally, the quadrupole does not seem to have a strong influence on $\Sigma$. For larger radii, we see the familiar behaviour of larger (smaller) values of $\Sigma$ for negative (positive) quadrupole $q$, similar to what we found for $T$, $P$, $F$ and $W$. As the ISCO is at smaller (larger) radial coordinates for negative (positive) quadrupole, it is clear that there must be a point where the difference to the Schwarzschild case is zero and then switches sign. This can be clearly seen in the difference plots \ref{sigmadifferenc01} and \ref{sigmadifferenc02}.

To  interpret  these  findings, it is tempting to look at the Newtonian picture.  We remember that a positive quadrupole $q>0$ can be interpreted as being due to a ring in the equatorial plane, whereas $q<0$ can correspond to two equal masses on the z-axis.  So $q > 0$ would cause a  net  force  directed  radially  outward,  and $q<0$  radially  inward.   The  relation  of  these  forces  to the position of the ISCO is shortly discussed in \cite{2016PhRvD..93f4019S}. The height scale result can also fit into this picture of Newtonian forces. However, for a consistent interpretation of our general results, including the up or down shifts and the different behaviour of quantities, the Newtonian setup seems to be too simplified, and further research in this direction is warranted.

\section{Acknowledgement}
We thank Petya Nedkova for very fruitful discussion. S.F. gratefully acknowledges useful discussions with Audrey Trova and Roberto Oliveri. The authors thank the research training group GRK 1620 ”Models of Gravity”, funded by the German Research Foundation (DFG). E.H. further acknowledges support from the DFG funded cluster of excellence Quantum Frontiers. 

\bibliographystyle{unsrt}
\bibliography{bibfirstmeeva}

\newcommand{\noop}[1]{}
  \providecommand{\noopsort}[1]{}\providecommand{\singleletter}[1]{#1}%
\begin{thebibliography}{10}

\bibitem{Abramowicz2013}
Marek~A. Abramowicz and P.~Chris Fragile.
\newblock Foundations of black hole accretion disk theory.
\newblock {\em Living Reviews in Relativity}, 16(1):1, Jan 2013.

\bibitem{1973A&A....24..337S}
N.~I. {Shakura} and R.~A. {Sunyaev}.
\newblock {Black holes in binary systems. Observational appearance.}
\newblock {\em \aap}, 24:337--355, 1973.

\bibitem{1972ApJ...178..347B}
J.~M. {Bardeen}, W.~H. {Press}, and S.~A. {Teukolsky}.
\newblock {Rotating Black Holes: Locally Nonrotating Frames, Energy Extraction,
  and Scalar Synchrotron Radiation}.
\newblock {\em \apj}, 178:347--370, December 1972.

\bibitem{1973blho.conf..343N}
I.~D. {Novikov} and K.~S. {Thorne}.
\newblock {Astrophysics of black holes.}
\newblock In C.~{Dewitt} and B.~S. {Dewitt}, editors, {\em Black Holes (Les
  Astres Occlus)}, pages 343--450, 1973.

\bibitem{1981AcA....31..283P}
B.~{Paczynski} and G.~{Bisnovatyi-Kogan}.
\newblock {A Model of a Thin Accretion Disk around a Black Hole}.
\newblock {\em \actaa}, 31:283, 1981.

\bibitem{1982AcA....32....1M}
B.~{Muchotrzeb} and B.~{Paczynski}.
\newblock {Transonic accretion flow in a thin disk around a black hole}.
\newblock {\em \actaa}, 32:1--11, 1982.

\bibitem{1988ApJ...332..646A}
M.~A. {Abramowicz}, B.~{Czerny}, J.~P. {Lasota}, and E.~{Szuszkiewicz}.
\newblock {Slim accretion disks}.
\newblock {\em \apj}, 332:646--658, September 1988.

\bibitem{1995ApJ...450..508R}
H.~{Riffert} and H.~{Herold}.
\newblock {Relativistic Accretion Disk Structure Revisited}.
\newblock {\em \apj}, 450:508, September 1995.

\bibitem{Stuchlik2004}
Z.~{Stuchl{\'{\i}}k} and P.~{Slan{\'y}}.
\newblock {Accretion disks in the Kerr-de Sitter spacetimes}.
\newblock In S.~{Hled{\'{\i}}k} and Z.~{Stuchl{\'{\i}}k}, editors, {\em RAGtime
  4/5: Workshops on black holes and neutron stars}, pages 205--237, December
  2004.

\bibitem{Chen2012}
Songbai {Chen} and Jiliang {Jing}.
\newblock {Properties of a thin accretion disk around a rotating non-Kerr black
  hole}.
\newblock {\em Physics Letters B}, 711(1):81--87, May 2012.

\bibitem{Kovacs2010}
Z.~{Kov{\'a}cs} and T.~{Harko}.
\newblock {Can accretion disk properties observationally distinguish black
  holes from naked singularities?}
\newblock {\em \prd}, 82(12):124047, Dec 2010.

\bibitem{Torres2002}
Diego~F. {Torres}.
\newblock {Accretion disc onto a static non-baryonic compact object}.
\newblock {\em Nuclear Physics B}, 626(1-2):377--394, Apr 2002.

\bibitem{Harko2009}
Tiberiu {Harko}, Zolt{\'a}n {Kov{\'a}cs}, and Francisco S.~N. {Lobo}.
\newblock {Thin accretion disks in stationary axisymmetric wormhole
  spacetimes}.
\newblock {\em \prd}, 79(6):064001, Mar 2009.

\bibitem{Kovacs2009}
Z.~{Kov{\'a}cs}, K.~S. {Cheng}, and T.~{Harko}.
\newblock {Thin accretion discs around neutron and quark stars}.
\newblock {\em \aap}, 500(2):621--631, Jun 2009.

\bibitem{Harko2009c}
Tiberiu {Harko}, Zolt{\'a}n {Kov{\'a}cs}, and Francisco S.~N. {Lobo}.
\newblock {Can accretion disk properties distinguish gravastars from black
  holes?}
\newblock {\em Classical and Quantum Gravity}, 26(21):215006, Nov 2009.

\bibitem{Danila2015}
Bogdan {D{\u{a}}nil{\u{a}}}, Tiberiu {Harko}, and Zolt{\'a}n {Kov{\'a}cs}.
\newblock {Thin accretion disks around cold Bose-Einstein condensate stars}.
\newblock {\em European Physical Journal C}, 75:203, May 2015.

\bibitem{Pun2008}
C.~S.~J. {Pun}, Z.~{Kov{\'a}cs}, and T.~{Harko}.
\newblock {Thin accretion disks in f(R) modified gravity models}.
\newblock {\em \prd}, 78(2):024043, Jul 2008.

\bibitem{Perez2013}
D.~{P{\'e}rez}, G.~E. {Romero}, and S.~E. {Perez Bergliaffa}.
\newblock {Accretion disks around black holes in modified strong gravity}.
\newblock {\em \aap}, 551:A4, Mar 2013.

\bibitem{Gyulchev2019}
Galin {Gyulchev}, Petya {Nedkova}, Tsvetan {Vetsov}, and Stoytcho {Yazadjiev}.
\newblock {Image of the Janis-Newman-Winicour naked singularity with a thin
  accretion disk}.
\newblock {\em \prd}, 100(2):024055, Jul 2019.

\bibitem{Harko2011}
Tiberiu {Harko}, Zolt{\'a}n {Kov{\'a}cs}, and Francisco S.~N. {Lobo}.
\newblock {Thin accretion disk signatures of slowly rotating black holes in
  Ho{\v{r}}ava gravity}.
\newblock {\em Classical and Quantum Gravity}, 28(16):165001, Aug 2011.

\bibitem{Heydari-Fard2010}
Malihe {Heydari-Fard}.
\newblock {Black hole accretion disks in brane gravity via a confining
  potential}.
\newblock {\em Classical and Quantum Gravity}, 27(23):235004, Dec 2010.

\bibitem{Harko2010}
Tiberiu {Harko}, Zolt{\'a}n {Kov{\'a}cs}, and Francisco S.~N. {Lobo}.
\newblock {Thin accretion disk signatures in dynamical Chern-Simons-modified
  gravity}.
\newblock {\em Classical and Quantum Gravity}, 27(10):105010, May 2010.

\bibitem{Perez2017}
Daniela {P{\'e}rez}, Federico G.~Lopez {Armengol}, and Gustavo~E. {Romero}.
\newblock {Accretion disks around black holes in scalar-tensor-vector gravity}.
\newblock {\em \prd}, 95(10):104047, May 2017.

\bibitem{Karimov2018}
R.~Kh. {Karimov}, R.~N. {Izmailov}, Amrita {Bhattacharya}, and K.~K. {Nandi}.
\newblock {Accretion disks around the
  Gibbons-Maeda-Garfinkle-Horowitz-Strominger charged black holes}.
\newblock {\em European Physical Journal C}, 78(9):788, Sep 2018.

\bibitem{Lodato2007}
G.~{Lodato}.
\newblock {Self-gravitating accretion discs}.
\newblock {\em Nuovo Cimento Rivista Serie}, 30:293, Jan 2007.

\bibitem{Will1974}
Clifford~M. {Will}.
\newblock {Perturbation of a Slowly Rotating Black Hole by a Stationary
  Axisymmetric Ring of Matter. I. Equilibrium Configurations}.
\newblock {\em \apj}, 191:521--532, Jul 1974.

\bibitem{Lemos1994}
Jos{\'e} P.~S. {Lemos} and Patricio~S. {Letelier}.
\newblock {Exact general relativistic thin disks around black holes}.
\newblock {\em \prd}, 49(10):5135--5143, May 1994.

\bibitem{1965ZhETF...49.170D}
A.~G. {Doroshkevich}, Y.~B. {Zel'dovich}, and I.~D. {Novikov}.
\newblock {Gravitational Collapse of Non-Symmetric and Rotating Bodies}.
\newblock {\em Zhurnal Eksperimentalnoi i Teoreticheskoi Fiziki}, 4:170,
  December 1965.

\bibitem{1982JMP....23..680G}
R.~{Geroch} and J.~B. {Hartle}.
\newblock {Distorted black holes.}
\newblock {\em Journal of Mathematical Physics}, 23:680--692, 1982.

\bibitem{Chandrasekhar:579245}
S~Chandrasekhar.
\newblock {\em {The mathematical theory of black holes}}.
\newblock Oxford classic texts in the physical sciences. Oxford Univ. Press,
  Oxford, 2002.

\bibitem{2016PhRvD..93f4019S}
A.~A. {Shoom}, C.~{Walsh}, and I.~{Booth}.
\newblock {Geodesic motion around a distorted static black hole}.
\newblock {\em \prd}, 93(6):064019, March 2016.

\bibitem{Grover2018}
Jai {Grover}, Jutta {Kunz}, Petya {Nedkova}, Alexand~er {Wittig}, and Stoytcho
  {Yazadjiev}.
\newblock {Multiple shadows from distorted static black holes}.
\newblock {\em \prd}, 97(8):084024, Apr 2018.

\bibitem{1997PhLA..230....7B}
Nora {Bret{\'o}n}, Tatiana~E. {Denisova}, and Vladimir~S. {Manko}.
\newblock {A Kerr black hole in the external gravitational field}.
\newblock {\em Physics Letters A}, 230:7--11, Feb 1997.

\bibitem{1990ForPh..38..733Q}
H.~{Quevedo}.
\newblock {Multipole Moments in General Relativity Static and Stationary Vacuum
  Solutions}.
\newblock {\em Fortschritte der Physik}, 38:733--840, 1990.

\bibitem{1974ApJ...191..499P}
D.~N. {Page} and K.~S. {Thorne}.
\newblock {Disk-Accretion onto a Black Hole. Time-Averaged Structure of
  Accretion Disk}.
\newblock {\em \apj}, 191:499--506, July 1974.

\bibitem{2012MNRAS.420..684P}
R.~F. {Penna}, A.~{S{\c a}owski}, and J.~C. {McKinney}.
\newblock {Thin-disc theory with a non-zero-torque boundary condition and
  comparisons with simulations}.
\newblock {\em \mnras}, 420:684--698, February 2012.

\bibitem{2017MNRAS.468.4351C}
G.~{Comp{\`e}re} and R.~{Oliveri}.
\newblock {Self-similar accretion in thin discs around near-extremal black
  holes}.
\newblock {\em \mnras}, 468:4351--4361, July 2017.

\bibitem{1997ApJ...479..179A}
M.~A. {Abramowicz}, A.~{Lanza}, and M.~J. {Percival}.
\newblock {Accretion Disks around Kerr Black Holes: Vertical Equilibrium
  Revisited}.
\newblock {\em \apj}, 479:179--183, April 1997.

\end{thebibliography}

\end{document}